# Electron doping and correlation effects on crystal, electronic and magnetic structures of $A_2NRuO_6$ ($A_2$ = $Ba_2$, BaLa; N = V, Cr, Fe)

M. Musa Saad H.-E.

Department of Physics, College of Science and Arts in Muthnib, Qassim University, 51931, Saudi Arabia

Fax: +966163421581, E-mail: 141261@qu.edu.sa

**Abstract**

Density functional methods have been used to study the crystal, electronic and magnetic structures of new ordered double perovskites $A_2NRuO_6$ ($A_2$ = $Ba_2$, BaLa; N = V, Cr, Fe). In the doped compounds, the A-site was replaced by 1:1 of Ba and La cations, $BaLaNRuO_6$. All compounds of $A_2NRuO_6$ crystallize in cubic symmetry with (space group Fm-3m and tilt system $a^0a^0a^0$). The electronic and magnetic calculations were performed by the full-potential linear muffin-tin orbital (PF-LMTO) method using both the generalized gradient approximation (GGA) and GGA plus on-site electron correlation effect (GGA+U). In GGA, $Ba_2NRuO_6$ shows half-metallic (HM), semiconducting and metallic behavior for N = V, Cr, Fe, respectively, completely change to HM when $A_2$ = BaLa. While, GGA+U method yields clearly HM state in all compounds, except for $Ba_2FeRuO_6$, shows an insulating behavior. Also, the results of magnetic structures calculations reveal that the $A_2NRuO_6$ compounds have ferrimagnetic (FI) nature if N = V and Cr, switch to ferromagnetic (FM) nature when N = Fe. The $V^{3+}$, $Cr^{3+}$, $Fe^{3+}$ and $Ru^{5+}$ ions are in high spin magnetic moments states due to the antiferromagnetic coupling N (3d) – O (2p) – Ru (4d).

## 1. Introduction

Half-metallic (HM) transition-metal double perovskites are extremely desired inorganic materials for spintronics applications [1], solid-state fuel cells (SOFC) design [2], semiconductor technology [3], as only the charge carriers having one of two possible spin polarization directions, spin-up ($n_\uparrow$) or spin-down ($n_\downarrow$), to transfer. The relative spin polarization ($S_P$) can be estimated using the ratio of the spin-density difference to the total spin-density at the Fermi level ($E_F$) as, $S_P = (n_\uparrow - n_\downarrow)/(n_\uparrow + n_\downarrow)$. The value of $S_P$ lie in the range $-1 \leq S_P \leq +1$, thus, $S_P = 0$ for un-polarized compounds, such as in semiconductors and insulators double perovskites, and $S_P = -1$ and $S_P = +1$ for the spin-down and spin-up full polarized compounds, as in half-metallic double perovskites [4-6]. Therefore, HM double perovskites have attracted special attention in many applied and fundamental area of solid-state physics, solid-state chemistry, materials science and materials engineering. As a result of these efforts, many different chemical and physical properties were observed in these materials. For example, HM combined with ferrimagnetic (FI) properties were detected in a few Sr-based compounds, $Sr_2CrMoO_6$ [4], $Sr_2CrWO_6$ [5] and $Sr_2FeMoO_6$ [6]. Based on electronic density of states (DOS) around the $E_F$, HM double perovskite shows a metallic nature with continuous DOS in spin-up or spin-down direction, while it is insulator or semiconductor with an energy-gap ($E_g$) in DOS for the other spin direction. HM nature is directly associated with the colossal magnetoresistance (MR) phenomena that revealed in some double perovskites, such as in $Sr_2MMoO_6$ (M = Cr) [4] and (M = Fe) [7].

Furthermore, many studies have been devoted to both the variation of the metallic and magnetic ions on N and M sites in transition-metal double perovskites $A_2NMO_6$ to modify the crystal, electronic and magnetic structures. These studies intended for optimizing the magnetic structures of double perovskites for many applications in magneto-electronic devices such as magnetic information storage systems, spin valves and spin polarization sources for spintronics devices. Along this way, important aspects are the achievement of sufficiently high Curie temperature ($T_C$) and spin polarization, such $T_C = 635$ K in HM $Sr_2CrReO_6$ [8], to allow for the operation of potential devices at room temperature (RT). Moreover, the increase of $T_C$ in $Sr_2FeMoO_6$ has been reported as a result of an electron doping by partially substitution of Sr site by La in $Sr_{2-x}La_xFeMoO_6$ [9]. With respect to MR, the large low field MR effect has been established in pure double perovskites, such $Sr_2CrWO_6$ [5], $Sr_2FeMoO_6$ [6,7] and in doped double perovskites, such $Sr_{2-x}La_xFeReO_6$ [10].

In this study, we performed first-principles calculations using the generalized gradient approximation (GGA) and GGA plus the exchange-correlation energy (GGA+U) to study the crystal, electronic and magnetic structures of six ordered double perovskites with the common chemical formula $A_2NRuO_6$ ($A_2$ = $Ba_2$, BaLa; N = V, Cr, Fe).

Moreover, we discussed the in detail effects of 1:1 electron doping of $A^{2+}$-site by $La^{3+}$, N-site substitution and exchange-correlation energy (GGA+U) on $A_2NRuO_6$ compounds. To the best of our knowledge, there are no experimental or theoretical reports on these compounds in literature, so, this is the first time to investigate this series in detailed. The crystal structure calculations showed that all these compounds have a cubic symmetry (Fm-3m). The rest of this paper is organized as follow; Section 2 gives a brief description of the computational methods and details of calculations used in this study, in the framework of the LMTART code. In Section 3, the results and discussion of the crystal, electronic and magnetic structures using the GGA and GGA+U calculations are present. Finally, the main conclusions are summarized in Section 4.

## 2. Calculation details

First-principles calculations in this study are performed by using the full potential linear muffin-tin orbital (FP-LMTO) within the atomic plane-wave (APLW) method [11] as implemented in LMTART code [12]. LMTART is designed to perform band structure, total energy and force calculations of solids using methods of density functional theory (DFT) [13]. In FP-LMTO technique, there is no shape approximation of the crystal potential; the crystal space is divided up into two main regions. Inside non-overlapping muffin-tin spheres (MTSs) surrounding all the atomic sites ($r \leq S$), where the potential V (r) is assumed to be spherically symmetric, and an interstitial region (IR), ($r > S$) separated the MTSs. The potential in IR is assumed to be constant $V_{MTZ}$ (r). Inside MTSs, the basis set is defined by the solution of radial Schrödinger equation of fixed energy multiplied by spherical harmonics [11]. While, in IR regions, the basis set consists of APLW are described by the Hankel function of kinetic energy $\kappa^2$ [12].

In order to describe correctly the PLW in IR, the spherical harmonics have been expanded up to 6.0 for all MTSs. The Brillouin zone (BZ) integration in the course of the self-consistency (SCF) iterations was performed over a *k*-point mesh of 6×6×6 with 120 *k*-points in the irreducible part of BZ. Double *κ* for (*s, p, d*) linear muffin-tin orbital basis is used, each radial-function inside the spheres matches to the HF in IR, for describing the valence states. The electronic configurations of elements in double perovskites $A_2NRuO_6$ ($A_2$ = $Ba_2$, BaLa; N = V, Cr, Fe) are Ba: [Xe] $6s^2$, La: [Xe] $5d^1$ $6s^2$, V: [Ar] $3d^3$ $4s^2$, Cr: [Ar] $3d^5$ $4s^1$, Fe: [Ar] $3d^6$ $4s^2$, Ru: [Kr] $4d^7$ $5s^1$, and O: [He] $2s^2$ $2p^4$. Therefore, the basis set consists of Ba (6s), La (6s), N (3d 4s), Ru (4d 5s) and O (2p) linear muffin-tin orbitals are taken as valence states, while the remains states are set as a frozen semi core states.

The generalized gradient approximation (GGA) in the framework of DFT is employed, where the spin polarization is unlimited to solve both spin-up and spin-down alignments. Therefore, GGA for the exchange-correlation energy ($E_{XC}$) in spin polarized system employs two spin densities $S_\uparrow$ and $S_\downarrow$, with total spin density of $S = S_\uparrow + S_\downarrow$. The $E_{XC}$ potential parameterization of Perdew-Wang 1991 (PW91) version of GGA was selected [11-13]. Furthermore, to treat strong electron-electron interaction effect in N (3d) and Ru (4d), the GGA+U calculations were carried out in the Hubbard model [12], via the option of Coulomb repulsion and the exchange energies (U = 4.0 eV, J = 0.89 eV) for N (3d) states and (U = 1.0 eV, J = 0.89 eV) for Ru (4d) states [4,14]. These exchange-correlation parameters were set using the Slater integrals $F^k$ (k ≥ 0). For d-electrons, k = 0, 2 and 4, therefore, only $F^0$, $F^2$ and $F^4$ are needed. For 3d and 4d states, they can be determined by the relationships $U = F^0$, $J = F^2+F^4/14$ and $F^2/F^4 = 5/6$.

## 3. Results and discussion

### 3.1 Crystal structures

Among oxides, double perovskites possess very flexible crystal structure $A_2NMO_6$, where A, M and N sites can be varied leading to a large number of compounds. Most double perovskites are distorted from the cubic structure; three factors are responsible for this distortion; size effect, deviations from ideal composition and Jahn-Teller effect. For size effect, according to the ionic radii, the tolerance factor allows to estimate the degree of distortion in $A_2NMO_6$, $TF = (\langle r_A \rangle + r_O)/\sqrt{2}(\langle r_{N,M} \rangle + r_O)$, where, $\langle r_A \rangle$, $\langle r_{N,M} \rangle$ and $\langle r_O \rangle$ are the average ionic radii of $A^{2+}$, $N^{3+}$–$M^{5+}$ and $O^{2-}$. Accordingly, cubic double perovskites have ($t \approx 1.00$); there is no tilting in $NO_6$ and $MO_6$ octahedra in order to fill the space in the crystal. Usually, the cubic structure of double perovskites occurs when (0.99 < t < 1.05) [15,16]. In this study, for $A_2NRuO_6$ ($A_2$ = $Ba_2$, BaLa; N = V, Cr, Fe), the effective ionic radii are $r(Ba^{2+})$ = 1.61 Å, $r(La^{3+})$ = 1.50 Å, $r(V^{3+})$ = 0.64 Å, $r(Cr^{3+})$ = 0.615 Å, $r(Fe^{3+})$ = 0.645 Å, $r(Ru^{5+})$ = 0.565 Å and $r(O^{2-})$ = 1.4 Å [16]. Therefore, the tolerance factors of these compounds were calculated and found to be amid in the range of ideal values, Table 1.

The structural information has been calculated using the SPuDS (Structure Prediction Diagnostic Software) [15], which allows the formula $A_2NMO_6$ in rock-salt cation ordering. The crystal symmetry, space-group (SG), Glazer tilt system (GTS), tolerance factor (TF), unit cell volume (V), lattice constants (a = b = c), oxygen position O (x), and the average bond distances and bond angles were calculated and summarized in Tables 1 and 2. It is found that the all compounds of $A_2NRuO_6$ crystallize in a cubic symmetry with space group (Fm-3m; No. 225) having an ideal lattice constant of a ≈ 8.00 Å [15,17]. Table 3 displays the atomic positions in the cubic symmetry of $A_2NRuO_6$. These results show that the cubic crystal structures of $A_2NRuO_6$ have not deviated from the ideal double perovskites, see Fig. 1, since there is (i) no displacement of N and Ru from the center of $NO_6$ and $RuO_6$ octahedra, N–O and Ru–O bonds are approximately equal, (ii) no displacement of the $A_2$ cations from the cavity centers, (iii) no distortion of $NO_6$ and $RuO_6$ octahedral cages, O–N–O = 90º and O–Ru–O = 90º and (iv) no tilting of the $NO_6$ and $RuO_6$ octahedra, N–O–Ru = 180º.

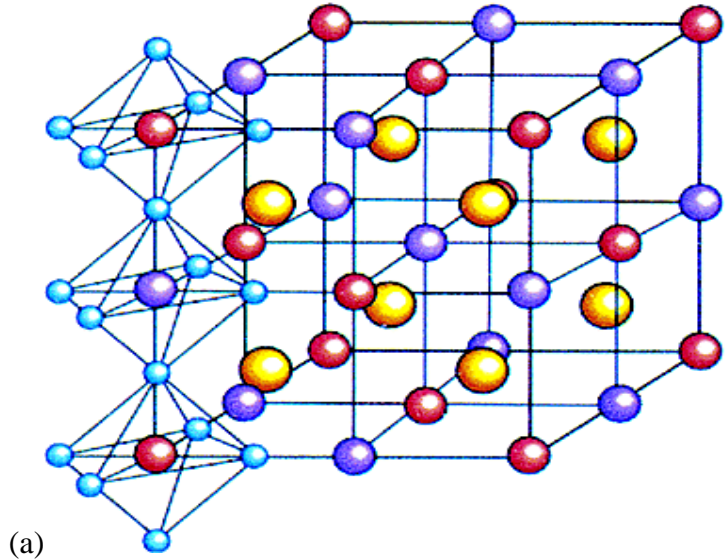

(a)

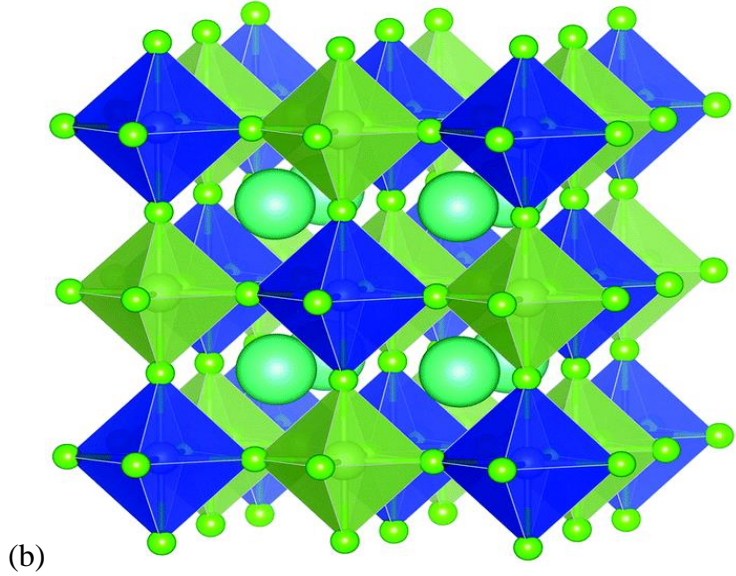

(b)

**Fig. 1.** 3D view of the crystal structure of double perovskite $A_2NMO_6$ with face centered cubic symmetry (space group of Fm-3m). The 1:1 ordering of N and M atoms are in the rock-salt arrangement. (a) The positions of atoms in unit cell, the yellow, red and violet spheres represent the A. N and M atoms, respectively. The oxygen atoms (cyan spheres) link the neighboring octahedra (b) the view of the corner-sharing octahedra $NO_6$ (blue octahedral) and $MO_6$ (green octahedral) that array in straight angle (N–O–M = 180º), where the A-cation (large green spheres) locating in the cavities formed by these octahedra.

**Table 1.** Calculated crystal structures data for the ordered double perovskites $A_2NRuO_6$ ($A_2$ = $Ba_2$, BaLa; N = V, Cr, Fe).

| $A_2NRuO_6$ | $A_2VRuO_6$ | | $A_2CrRuO_6$ | | $A_2FeRuO_6$ | |
|---|---|---|---|---|---|---|
| $A_2$-sites | $Ba_2$ | BaLa | $Ba_2$ | BaLa | $Ba_2$ | BaLa |
| Symmetry | FCC | FCC | FCC | FCC | FCC | FCC |
| S.G. | Fm-3m | Fm-3m | Fm-3m | Fm-3m | Fm-3m | Fm-3m |
| GTS | $a^0a^0a^0$ | $a^0a^0a^0$ | $a^0a^0a^0$ | $a^0a^0a^0$ | $a^0a^0a^0$ | $a^0a^0a^0$ |
| TF | 1.0509 | 0.9999 | 1.0560 | 1.0047 | 1.0467 | 0.9959 |
| V (Å$^3$) | 499.403 | 505.685 | 492.262 | 498.484 | 505.470 | 511.803 |
| a = b = c (Å) | 7.9338 | 7.9670 | 7.8958 | 7.9290 | 7.9658 | 7.9990 |
| O (x) | 0.2520 | 0.2510 | 0.2508 | 0.2498 | 0.2530 | 0.2520 |

**Table 2.** Calculated bond distances and angles in the cubic crystals of ordered double perovskites $A_2NRuO_6$ ($A_2$ = $Ba_2$, BaLa; N = V, Cr, Fe).

| $A_2NRuO_6$ | $A_2VRuO_6$ | | $A_2CrRuO_6$ | | $A_2FeRuO_6$ | |
|---|---|---|---|---|---|---|
| $A_2$-sites | $Ba_2$ | BaLa | $Ba_2$ | BaLa | $Ba_2$ | BaLa |
| **Bond distances** | | | | | | |
| La (8c)–O (24e) ×12 (Å) | | 2.8168 | | 2.8033 | | 2.8281 |
| Ba (8c)–O (24e) ×12 (Å) | 2.8051 | 2.8168 | 2.7916 | 2.8033 | 2.8165 | 2.8281 |
| N (4a)–O (24e) ×6 (Å) | 1.9995 | 1.9995 | 1.9805 | 1.9804 | 2.0155 | 2.0155 |
| Ru (4b)–O (24e) ×6 (Å) | 1.9675 | 1.9840 | 1.9675 | 1.9840 | 1.9675 | 1.9840 |
| **Bond angles** | | | | | | |
| O (24e)–N (4a)–O (24e) (°) | 90.00 | 90.00 | 90.00 | 90.00 | 90.00 | 90.00 |
| O (24e)–Ru (4b)–O (24e) (°) | 90.00 | 90.00 | 90.00 | 90.00 | 90.00 | 90.00 |
| N (4a)–O (24e)–Ru (4b) (°) | 180.0 | 180.0 | 180.0 | 180.0 | 180.0 | 180.0 |

**Table 3.** Atoms, multiplicity, Wyckoff, symmetry, positions and occupancy for the ordered double perovskites $A_2NRuO_6$ ($A_2$ = $Ba_2$, BaLa; N = V, Cr, Fe). The coordinate values of O (x) are listed in Table 1.

| Atom | Multiplicity | Wyckoff | Symmetry | Positions | | | Occupancy |
|---|---|---|---|---|---|---|---|
| | | | | x | y | z | |
| Ba | 8 | c | -43m | ¼ | ¼ | ¼ | 0.5 |
| La | 8 | c | -43m | ¼ | ¼ | ¼ | 0.5 |
| N | 4 | a | m-3m | 0 | 0 | 0 | 1.0 |
| Ru | 4 | b | m-3m | ½ | ½ | ½ | 1.0 |
| O | 24 | e | 4mm | x | 0 | 0 | 1.0 |

### 3.2 Electronic structures

The total density of states (TDOS) and partial density of states (PDOS) of ordered double perovskites $A_2NRuO_6$ ($A_2$ = $Ba_2$, BaLa; N = V, Cr, Fe) were calculated using the GGA and GGA+U methods. The electronic DFT-TDOS plots per unit cell for spin-up, upper curves (+TDOS) and spin-down, lower curves (–TDOS) are collectively presented in Fig. 2. The horizontal axis stands for the energy relative to the Fermi energy [E (eV)], thus, the Fermi level ($E_F$) is situated at zero energy (dash line; $E_F$ = 0.0 eV). Figs. 2–6 show the TDOS and PDOS behavior for these compounds, plotted between –8.0 eV and +8.0 eV, where the main electronic features occur. In order to realize the effects of electron doping, $A_2$ and N sites substitution, and the correlation energy, we discuss the electronic structures of $A_2NRuO_6$ in a general discussion. Figs. 2 and 3 show the combined GGA and GGA+U curves of the TDOS of these compounds. From the GGA TDOSs around the $E_F$, it can see that the $Ba_2NRuO_6$ compounds show different electronic structures; HM nature in N = V, semiconducting nature in N = Cr and metallic nature when N = Fe. These structures are completely transit to HM state when the $A_2$ sites are replaced by BaLa atoms. When the correlated GGA+U method is applied, a clearly HM electronic nature is maintained in all $A_2NRuO_6$ compounds, except for the N = Fe compound with $A_2$ = $Ba_2$, shows an electronic insulating state. The obtained results are in a good agreement with the pervious results [17,18,19]. Table 4 summarizes the results of the electronic energy-gap ($E_g$) and band-width (W) at the $E_F$, in spin-up and spin-down directions, for $A_2NRuO_6$ obtained from GGA and GGA+U calculations.

**Table 4.** Spin-up (↑) and spin-down (↓) energy-gap ($E_g$) and band-width (W) from the TDOS (Figs. 2 and 3) of six double perovskite oxides $A_2NRuO_6$ ($A_2$ = $Ba_2$, BaLa; N = V, Cr, Fe), calculated using GGA and GGA+U methods.

| $A_2NRuO_6$ | | $A_2VRuO_6$ | | $A_2CrRuO_6$ | | $A_2FeRuO_6$ | |
|---|---|---|---|---|---|---|---|
| $A_2$-sites | | $Ba_2$ | BaLa | $Ba_2$ | BaLa | $Ba_2$ | BaLa |
| GGA | $E_g\uparrow$ (eV) | 0.68 | 0.00 | 0.94 | 0.00 | 0.00 | 1.73 |
| | $E_g\downarrow$ (eV) | 0.00 | 0.77 | 0.14 | 0.66 | 0.00 | 0.00 |
| | $W\uparrow$ (eV) | 0.00 | 2.94 | 0.00 | 1.91 | 1.59 | 0.00 |
| | $W\downarrow$ (eV) | 2.43 | 0.00 | 0.00 | 0.00 | 1.81 | 2.53 |
| GGA+U | $E_g\uparrow$ (eV) | 1.89 | 0.00 | 2.24 | 0.00 | 1.82 | 1.61 |
| | $E_g\downarrow$ (eV) | 0.00 | 0.94 | 1.48 | 1.34 | 1.82 | 0.00 |
| | $W\uparrow$ (eV) | 0.00 | 2.40 | 0.00 | 0.93 | 0.00 | 0.00 |
| | $W\downarrow$ (eV) | 1.98 | 0.00 | 0.00 | 0.00 | 0.00 | 1.73 |

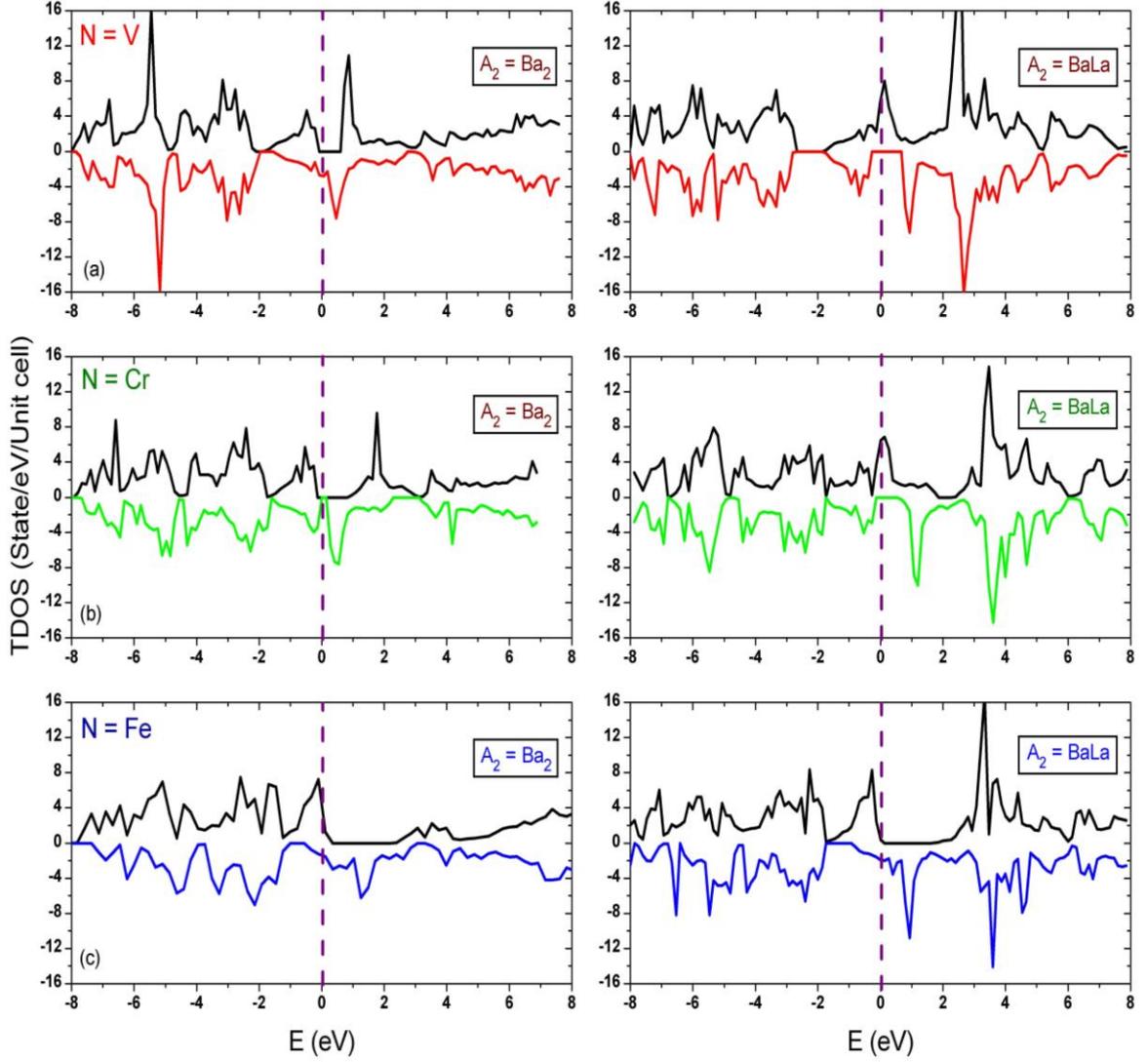

**Fig. 2.** The spin-up and spin-down GGA total density of states (TDOS) of double perovskites $A_2NRuO_6$, $A_2 = Ba_2$ (left panels), $A_2 =$ BaLa (right panels); (a) N = V, (b) N = Cr, and (c) N = Fe. Fermi level (dash line) was set at the zero energy ($E_F = 0.0$ eV).

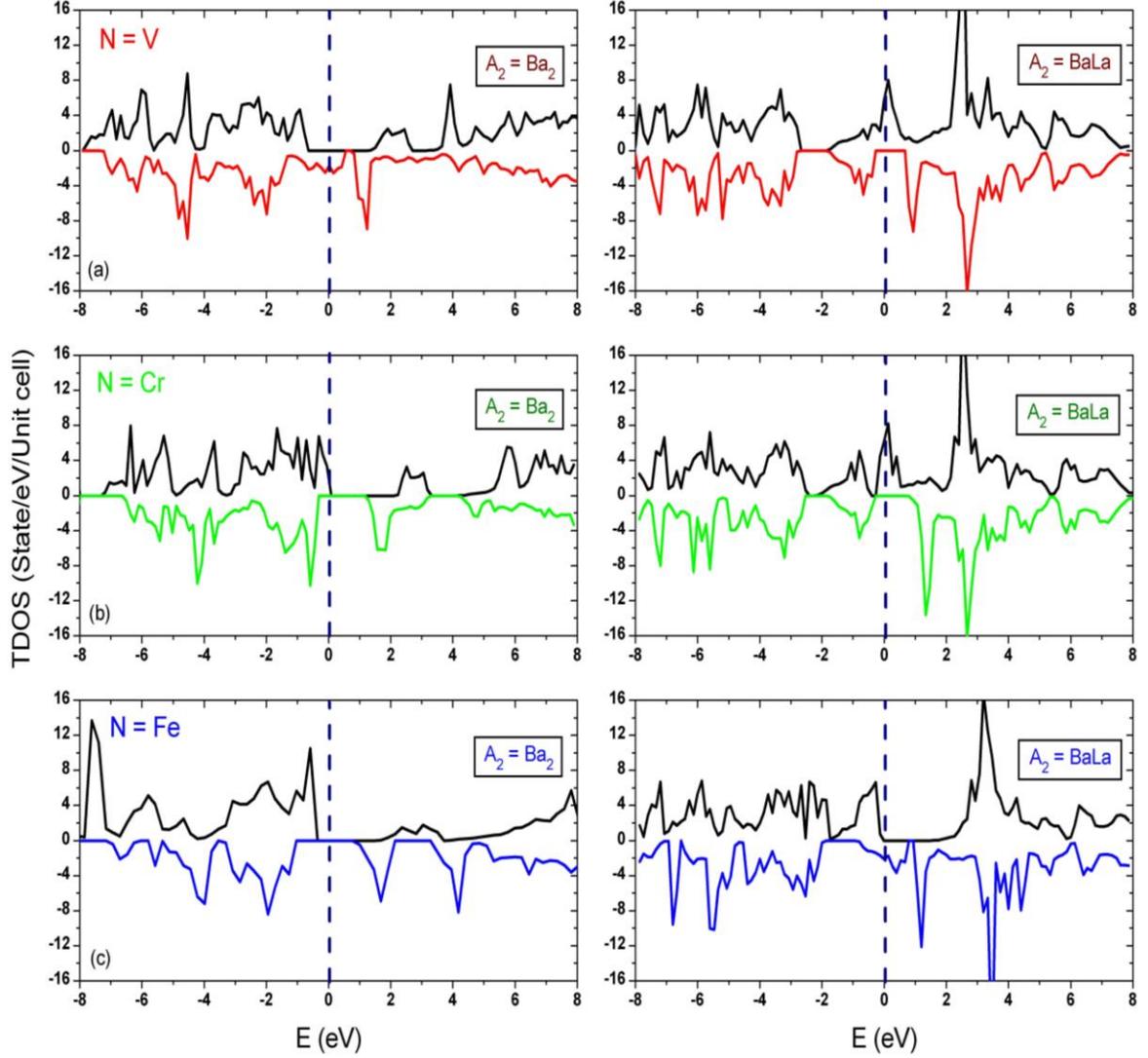

**Fig. 3.** The spin-up and spin-down GGA+U total density of states (TDOS) of double perovskites $A_2NRuO_6$, $A_2 = Ba_2$ (left panels), $A_2 = BaLa$ (right panels); (a) N = V, (b) N = Cr, and (c) N = Fe. Fermi level (dash line) was set at the zero energy ($E_F = 0.0$ eV).

To gain deep insight into the partial contributions to the TDOS and various phenomena, the projected partial density of states (PDOS) of M (3d), Ru (4d) and O (2p) are calculated using the GGA and GGA+U methods. Figs. 4, 5 and 6 show the spin-up and spin-down partial density of states (PDOS) of M (3d), Ru (4d) and O (2p) ions in six compounds. The PDOS of Ba and La ions are not given here since negligible interactions are observed between Ba and La ions and other ions in double perovskites $A_2NRuO_6$. From Fig. 4, we can see that the conduction band in spin-down between –1.5 and +1.5 eV almost come from the hybridization of V (3d) and O (2p) electrons in $Ba_2VRuO_6$. Switch to spin-up HM in $BaLaVRuO_6$ compound, here, the conduction band between –1.5 and +1.5 eV composes from the hybridization of Ru (4d) and O (2p) states. In semiconductor Ba$_2$CrRuO$_6$, Fig. 5, the spin-up and spin-down states of Cr (3d), Ru (4d) and O (2p) have significant contributions at the valence and conduction bands. Whereas, HM BaLaCrRuO$_6$ compound, the spin-up conduction band almost yields from the hybridizations of both Cr (3d)–O (2p) and Ru (4d) –O (2p) states between –1.5 and +1.5 eV. For the metallic compound Ba$_2$FeRuO$_6$, Fig. 6, the spin-up and spin-down conduction band contain all the partial PDOS of Fe (3d), Ru (4d) and O (2p) states in GGA. While, the spin-down HM in BaLaFeRuO$_6$ compound, the conduction band come from the hybridization of Fe (3d) and O (2p) states.

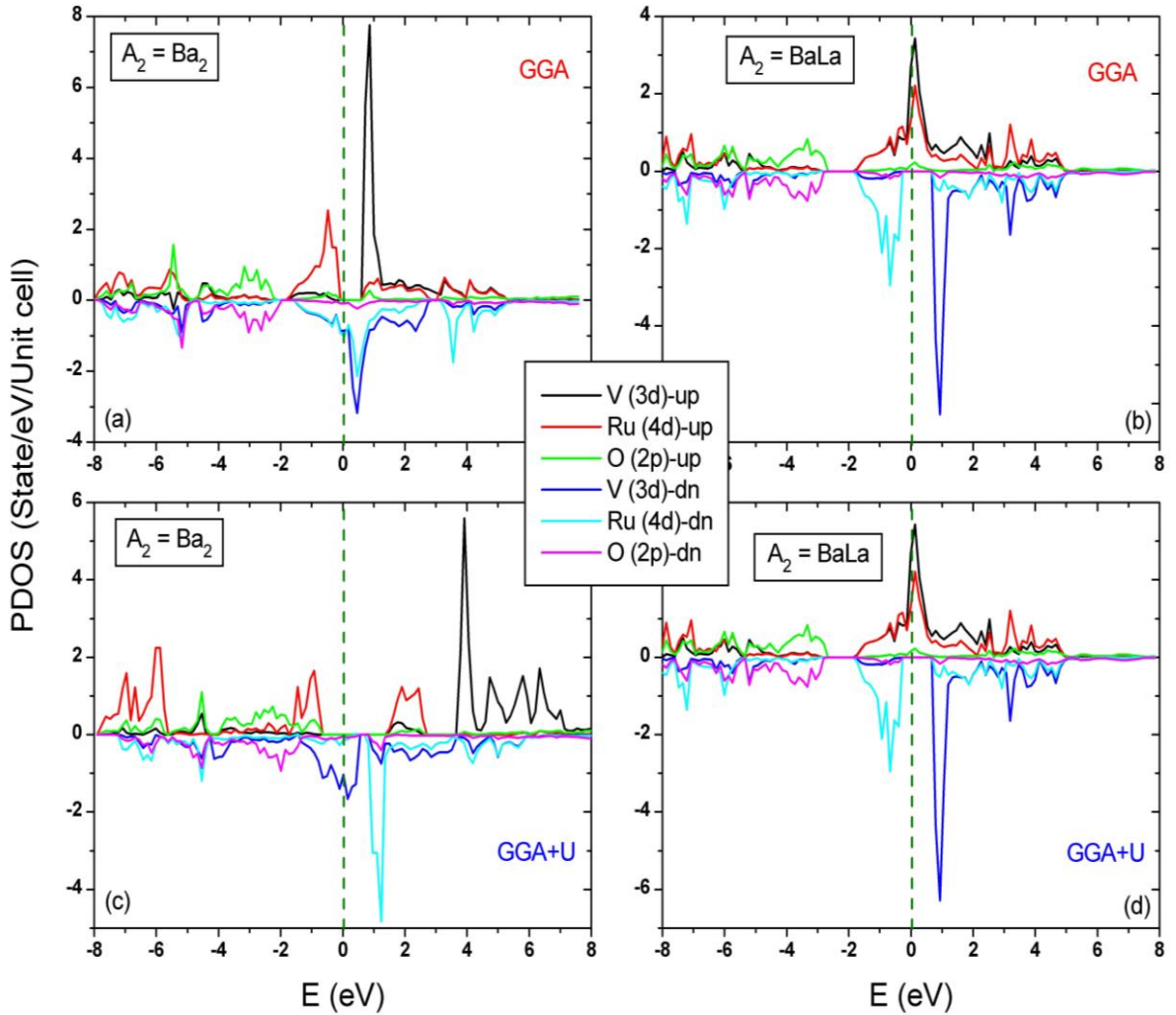

**Fig. 4.** The spin-up and spin-down partial density of states (PDOS) of V (3d), (c) Ru (4d) and (d) O (2p) states in double perovskites A$_2$VRuO$_6$, A$_2$ = Ba$_2$ (left panels), A$_2$ = BaLa (right panels), calculated by (a)-(b) GGA, and (c)-(d) GGA+U methods. Fermi level (dash line) was set at the zero energy (E$_F$ = 0.0 eV).

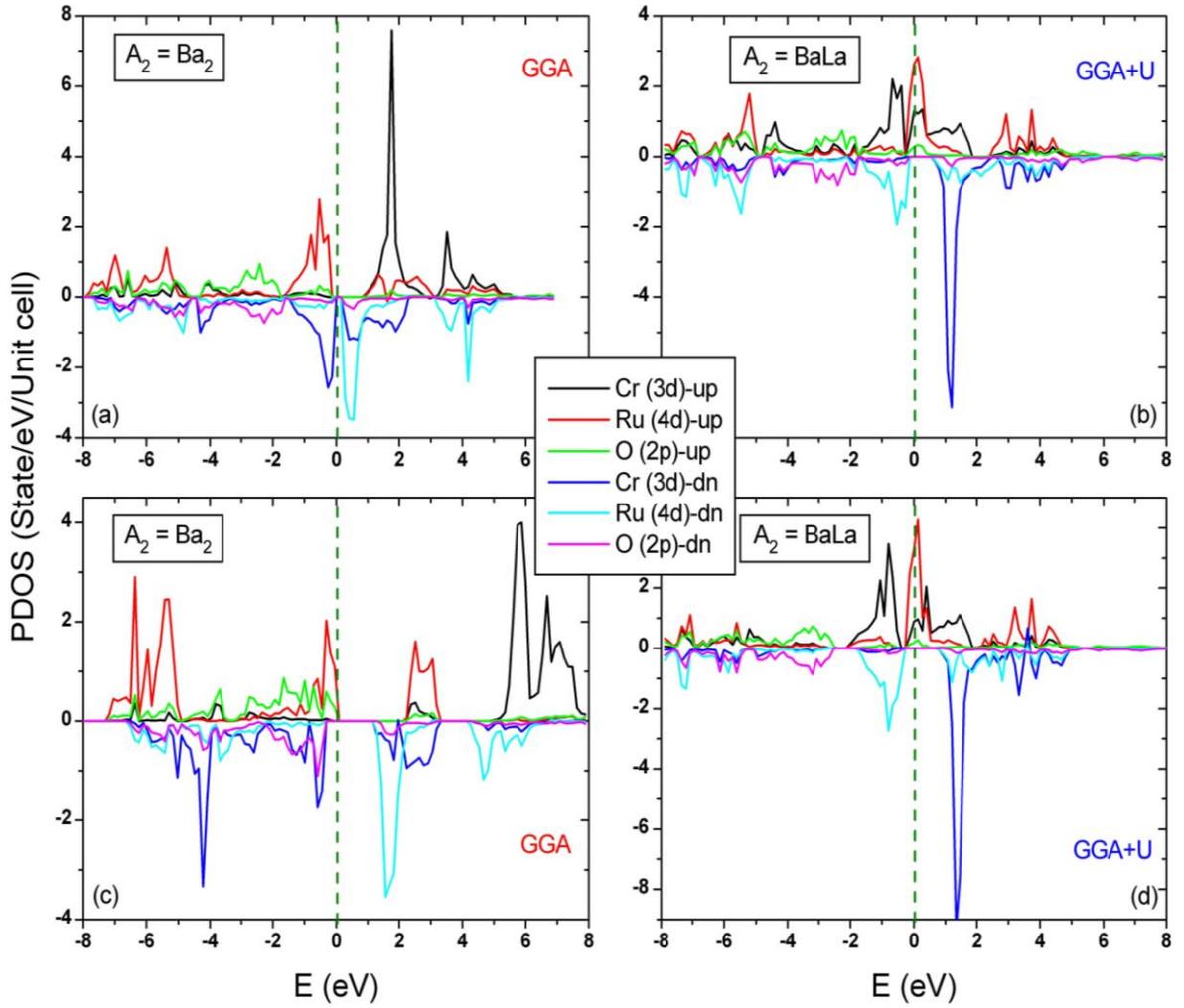

**Fig. 5.** The spin-up and spin-down partial density of states (PDOS) of Cr (3d), (c) Ru (4d) and (d) O (2p) states in double perovskites $A_2CrRuO_6$, $A_2$ = $Ba_2$ (left panels), $A_2$ = BaLa (right panels), calculated by (a)-(b) GGA, and (c)-(d) GGA+U methods. Fermi level (dash line) was set at the zero energy ($E_F$ = 0.0 eV).

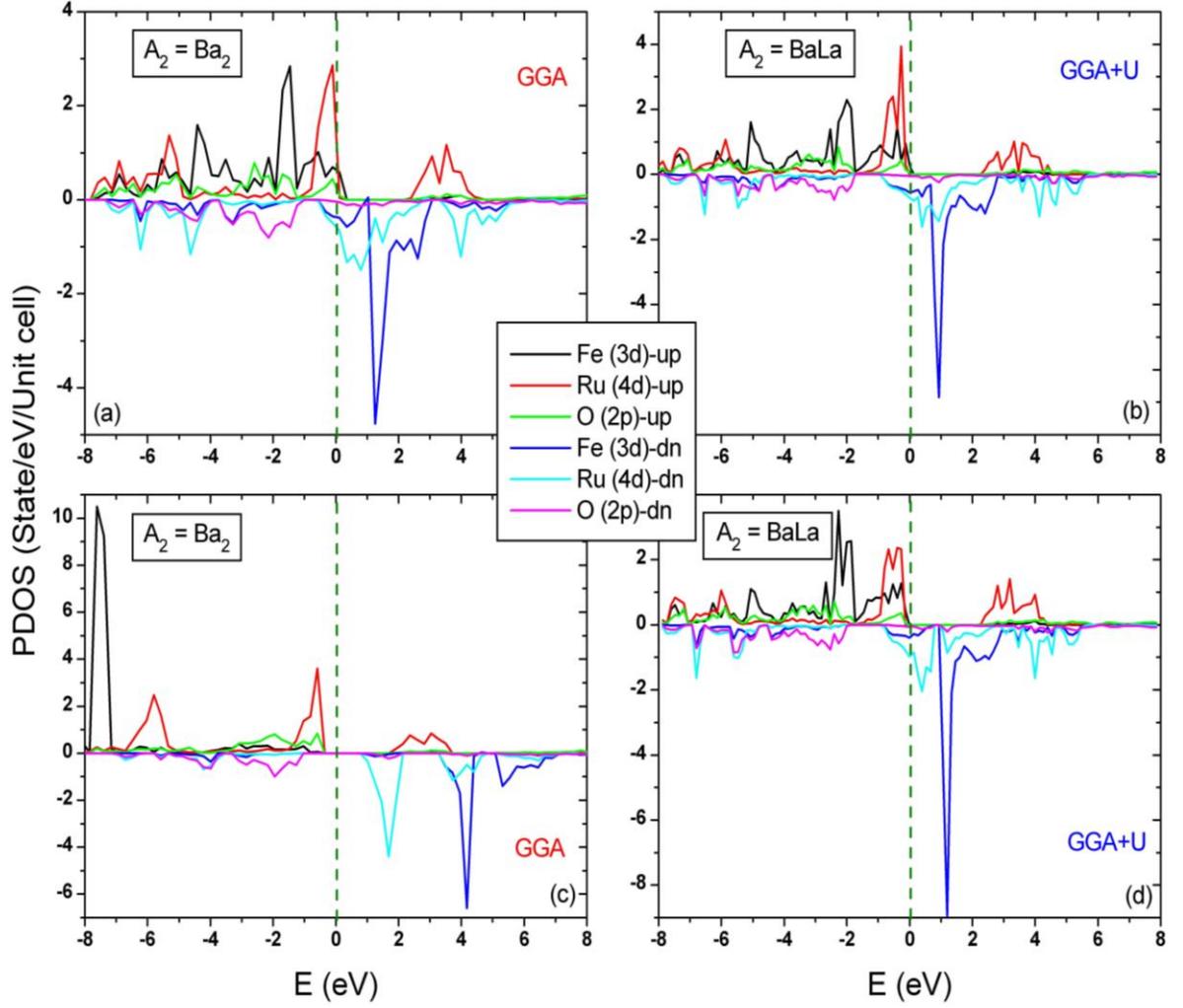

**Fig. 6.** The spin-up and spin-down partial density of states (PDOS) of Fe (3d), (c) Ru (4d) and (d) O (2p) states in double perovskites $A_2FeRuO_6$, $A_2$ = $Ba_2$ (left panels), $A_2$ = BaLa (right panels), calculated by (a)-(b) GGA, and (c)-(d) GGA+U methods. Fermi level (dash line) was set at the zero energy ($E_F$ = 0.0 eV).

### 3.3 Magnetic structures

To study the magnetic structures of double perovskite oxides $A_2NRuO_6$ ($A_2$ = $Ba_2$, BaLa; N = V, Cr, Fe), the spin-polarized DFT calculations using the GGA and GGA+U methods were carried out. The results of these calculations reveal that the N = V and N = Cr compounds possess ferrimagnetic (FI) nature within two methods, while, the N = Fe compounds possess ferromagnetic (FM) nature. The major source of the magnetization in $A_2NRuO_6$ compounds is the $N^{3+}$ (3d) and $Ru^{5+}$ (4d) orbitals. Further analysis of the $N^{3+}$ (3d) states shows that the 3d-$t_{2g}$ orbitals are partially filled and the 3d-$e_g$ orbitals are empty or filled by 2 electrons, as in N = Fe. This reveals that the electronic configurations of $N^{3+}$ (3d) ions are $V^{3+}$ (3$d^2$; $t_{2g}^2\uparrow t_{2g}^0\downarrow\ e_g^0\uparrow\ e_g^0\downarrow$; S = 2/2), $Cr^{3+}$ (3$d^3$; $t_{2g}^3\uparrow t_{2g}^0\downarrow\ e_g^0\uparrow\ e_g^0\downarrow$; S = 3/2), and $Fe^{3+}$ (3$d^5$; $t_{2g}^3\uparrow t_{2g}^0\downarrow\ e_g^2\uparrow\ e_g^0\downarrow$; S = 5/2). As a result, the $N^{3+}$ (3d) ions carry most of the spin magnetic moment, making them responsible for the FI and FM natures in three $A_2NRuO_6$ compounds. While in $Ru^{5+}$ (4d) states, the 4d-$t_{2g}$ orbitals are partially filled by three electrons and 4d-$e_g$ orbitals are empty; $Ru^{5+}$ (4$d^3$; $t_{2g}^3\uparrow t_{2g}^0\downarrow\ e_g^0\uparrow\ e_g^0\downarrow$; S = 3/2), thus, they hold a small spin magnetic moment that induced by the N (3d) – O (2p) – Ru (4d) hybridizations. The total spin magnetic moments per unit cell of these compounds, as well their partial spin magnetic moments per atom are calculated and summarized in Table 5.

**Table 5.** Calculated the partial and total spin magnetic moments per unit cell (in $\mu_B$ unit) of double perovskites $A_2NRuO_6$ ($A_2$ = $Ba_2$, BaLa; N = V, Cr, Fe) by GGA and GGA+U methods.

| $A_2NRuO_6$ | | $A_2VRuO_6$ | | $A_2CrRuO_6$ | | $A_2FeRuO_6$ | |
|---|---|---|---|---|---|---|---|
| $A_2$-sites | | $Ba_2$ | BaLa | $Ba_2$ | BaLa | $Ba_2$ | BaLa |
| GGA | Ba | 0.042 | -0.004 | 0.056 | -0.029 | 0.062 | 0.055 |
| | La | - | 0.018 | - | 0.006 | - | 0.085 |
| | N | -0.744 | 1.051 | -2.123 | 2.040 | 4.213 | 3.729 |
| | Ru | 1.139 | -0.818 | 1.447 | -0.760 | 2.203 | 1.676 |
| | O | 0.488 | -0.226 | 0.560 | -0.318 | 0.720 | 0.579 |
| GGA+U | Ba | 0.074 | -0.004 | 0.073 | -0.005 | 0.047 | 0.048 |
| | La | - | 0.018 | - | 0.024 | - | 0.074 |
| | N | -1.818 | 1.051 | -2.034 | 2.301 | 4.502 | 3.923 |
| | Ru | 1.896 | -0.818 | 2.034 | -1.050 | 2.361 | 1.622 |
| | O | 0.732 | -0.229 | 0.762 | -0.290 | 0.714 | 0.528 |

**4. Conclusions**

In this study, the crystal, electronic and magnetic structures of double perovskites $A_2NRuO_6$ ($A_2$ = $Ba_2$, BaLa; N = V, Cr, Fe) have been studied using the FP-LMTO method based on the DFT. Moreover, we employed the GGA and GGA+U methods. The crystal structures calculations showed that the $A_2NRuO_6$ crystallized in cubic symmetry with space group Fm-3m and tilt system $a^0a^0a^0$. The electronic structures calculations showed that the GGA and GGA+U gave different electronic phases; $Ba_2NRuO_6$ in GGA shows half-metallic (HM), semiconducting and metallic behavior for N = V, Cr, Fe, respectively, changed to HM if $A_2$ = BaLa. The GGA+U method produced HM state in all compounds, except for the $Ba_2FeRuO_6$, it showed an insulating behavior. The magnetic structures calculations revealed that the $A_2NRuO_6$ compounds have a ferrimagnetic (FI) nature if N = V and Cr, switched to a ferromagnetic (FM) nature when N = Fe. The analysis demonstrated that the $V^{3+}$, $Cr^{3+}$, $Fe^{3+}$ and $Ru^{5+}$ ions are in high spin magnetic moments states due to the antiferromagnetic coupling N (3d) – O (2p) – Ru (4d).